\pdfoutput=1
\documentclass[aps,prl,floatfix,superscriptaddress,twocolumn,reprint]{revtex4-1}
\usepackage{eurosym}
\usepackage{amsbsy}
\usepackage{latexsym,epsfig,graphicx}
\usepackage{dcolumn}
\usepackage{subfigure}
\usepackage{comment}
\usepackage{color}
\usepackage{bm}
\usepackage{mathrsfs}
\usepackage{amssymb}
\usepackage{amsfonts}
\usepackage{amsmath}
\usepackage{xspace}
\usepackage{epstopdf}
\usepackage{tabularx}
\usepackage{longtable}
\usepackage[colorlinks=true, letterpaper=true, pdfstartview=FitV, linkcolor=blue, citecolor=blue, urlcolor=blue]{hyperref}
\usepackage[normalem]{ulem}
\usepackage[version=4]{mhchem}

\newcommand{\sgn}{\text{sgn}}

\begin{document}
\title{Topological Invariants for Quantum Quench Dynamics from Unitary Evolution}
\author{Haiping Hu}
\affiliation{Department of Physics and Astronomy, George Mason University, Fairfax, Virginia 22030, USA}
\affiliation{Department of Physics and Astronomy, University of Pittsburgh, Pittsburgh, Pennsylvania 15260, USA}
\author{Erhai Zhao}
\email{ezhao2@gmu.edu}
\affiliation{Department of Physics and Astronomy, George Mason University, Fairfax, Virginia 22030, USA}
\date{\today}
\begin{abstract}
Recent experiments began to explore the topological properties of quench dynamics, i.e. the time evolution following a sudden change in the Hamiltonian, via tomography of quantum gases in optical lattices. In contrast to the well established theory for static band insulators or periodically driven systems, at present it is not clear whether, and how, topological invariants can be defined for a general quench of band insulators. Previous work solved a special case of this problem beautifully using Hopf mapping of two-band Hamiltonians in two dimensions. But it only works for topologically trivial initial state and is hard to generalize to multiband systems or other dimensions. Here we introduce the concept of loop unitary constructed from the unitary time-evolution operator, and show its homotopy invariant fully characterizes the dynamical topology. For two-band systems in two dimensions, we prove that the invariant is precisely equal to the change in the Chern number across the quench regardless of the initial state. We further show that the nontrivial dynamical topology manifests as hedgehog defects in the loop unitary, and also as winding and linking of its eigenvectors along a curve where dynamical quantum phase transition occurs. This opens up a systematic route to classify and characterize quantum quench dynamics. 
\end{abstract}
\maketitle

{\color{blue}\textit{Introduction.}} Topological structures are ubiquitous in nature. They appear either in real space, e.g. a smoke ring or linked coronal loops near the Sun, or in momentum space, e.g. skymions in a Chern insulator or hedgehogs in a Weyl semimetal. Recently,
the investigation of topological phenomena \cite{review1,review2} extended to time-dependent quantum systems, e.g. Floquet systems under periodic driving \cite{fl1,fl2,fl3,fl4,fl5,fl6,fl7,fl8,roy,wangzhong,zoulw} or quench dynamics following a sudden change in the Hamiltonian, $H_0\rightarrow H$ \cite{qq1,qq2,qq3,qq4,qq5,qq6,qq7,qq8,qq9,qq10,qq11,qq12,qq13,qq14,qq15,qq16,qq17}. For these dynamical systems, the topological structures are hidden in the momentum-time continuum. Intriguingly, there seems a deep connection between the static band topology and quench dynamics. For example, topological insulators in integer classes can be systematically classified by quantum quenches starting from a trivial state based on the dynamical bulk-surface correspondence \cite{qq8,qq13}. Quench dynamics has been measured in details for example in experiments on ultracold atoms \cite{quenchexp2,quenchexp3,quenchexp4} and photonic quantum walks \cite{xuep1,xuep2}. This raises the question, how to systematically construct the topological invariants for the quench dynamics of band insulators?

The answer to this question remains open. Previous work on the quench dynamics of two-band Bloch Hamiltonians cast the problem mathematically as a Hopf map to arrive at a powerful result: the time evolution is characterized by the so-called Hopf invariant, which counts the linking number of the preimages of two time-evolved states, and equals to the Chern number of the post-quench Hamiltonian $H$ \cite{qq5}. Experimentally, such Hopf links in the momentum-time space have been observed using Bloch-state tomography for ultracold $^{40}$K and $^{87}$Rb atoms in optical lattices \cite{quenchexp2,quenchexp3,quenchexp4}. This framework based on Hopf mapping however works only for two-band Hamiltonians in two dimensions (2D). Moreover, it requires the pre-quench Hamiltonian $H_0$ to be topologically trivial. For quenches from a nontrivial state, the Hopf invariant is no longer well-defined \cite{wilzeck,chenxin,csintegral}. Thus, a unified theory valid for general quench is still lacking. A proper topological invariant that remains well defined regardless of the triviality of the initial state is highly desired.

The quench dynamics of Chern insulators has also been studied from the perspective of dynamical quantum phase transition (DQPT) \cite{dqpt1,dqpt2,dqpt3,dqpt4,dqpt5,dqpt6,dqpt7,dqpt8,dqpt9,dqptexp1,dqptexp2,quenchexp1}. A DQPT is identified when physical observables show nonanalytic behavior at some time instant, e.g. when the post-quench state becomes orthogonal to the initial state \cite{dqpt3}. It remains unclear how DQPT is related to the topological invariants for quench dynamics. Relatedly, periodically driven systems have been systematically classified into the periodic table of Floquet topological insulators \cite{roy,wangzhong}. Possible connection between Floquet dynamics and quench dynamics, however, has not been noticed or emphasized.

To address these questions, we propose a new framework to characterize the topological properties of quench dynamics. We introduce the loop unitary $U_l$ and show its homotopy invariant $W_3$ relates the pre- and post-quench Chern numbers by $W_3=\mathcal{C}_f-\mathcal{C}_i$, which works for any $H_0$ and $H$. For trivial initial state, $W_3$ reduces to the Hopf invariant \cite{qq5}. We reveal the origin of dynamical topology in the hedgehogs ($\pi$-defects) of the phase band of $U_l$. Moreover, we introduce the notion of DQPT curve to show the dynamical topology also manifests as the windings, links, or knots of the eigenvector of $U_l$ along this curve. We illustrate our theory by applying it to a highly tunable model of two-band Hamiltonian in 2D. The framework paves the way to study the topological properties of more general quantum quenches.

{\color{blue}\textit{Loop unitary for quantum quench.}} We will focus on the quantum quench dynamics of a generic two-band system in 2D. The system starts from an arbitrary initial state $|\xi_0\rangle$ at time zero, then evolves according to a post-quench Hamiltonian $H=\bm H(\bm k)\cdot\bm\sigma$, where the quasimomentum $\bm k=(k_x,k_y)$ and $\bm \sigma=(\sigma_x,\sigma_y,\sigma_z)$ is the Pauli matrix. At a later time $\tau$, the state evolves into $|\xi(\tau)\rangle=e^{-iH\tau}|\xi_0\rangle$ with $\hbar=1$. Let $\pm E_{\bm k}$ be the spectrum of $H$, and assume $H$ is gapped, $E_{\bm k}>0$. As far as the topological properties are concerned, we can rescale $H$ and replace it with $h={H}/{E_{\bm k}}=\hat{\bm h}(\bm k)\cdot\bm\sigma$ where $\hat{\bm h}$ is a unit vector. This amounts to the standard band flattening. Equivalently, we can view ${\bm H}\rightarrow \hat{\bm h}$ as a rescaling of time $\tau\rightarrow t=E_{\bm k}\tau$. The rescaling yields a key observation: at $t=\pi$, the state returns to the initial state up to a minus sign, $|\xi(\pi)\rangle=-|\xi_0\rangle$. Thus, the quench dynamics has period $\pi$. 

Previous works represent the two-component spinor $|\xi(t)\rangle$ with a point $\bm\xi=\langle\xi|\bm\sigma|\xi\rangle$ on the Bloch sphere $S^2$. This defines a mapping from the $(\bm k,t)$-space, a three-torus $\textrm{T}^3$, to $S^2$ with homotopy group $\pi_3(S^2)=\mathbb{Z}$ \cite{hopf1,DLM,thesis}. If the initial state $|\xi_0\rangle$ is topologically trivial with Chern number $\mathcal{C}_i=0$, the quench dynamics can be characterized by the Hopf invariant as a Chern-Simons integral \cite{qq5}
\begin{eqnarray}
\mathcal{L}=\frac{1}{4\pi^2}\int_{\textrm{T}^3}d^2\bm kdt~\epsilon^{\mu\nu\rho}\mathcal{A}_{\mu}\partial_{\nu}\mathcal{A}_{\rho}.
\end{eqnarray}
Here $\mathcal{A}_{\mu}=i\langle\xi |\partial_{\mu}|\xi\rangle$ is the Berry connection, the indices $(\mu\nu\rho)$ take values in $(k_x, k_y, t)$, and Einstein's summation convention is used. It is proved in Ref. \cite{qq5} that $\mathcal{L}=\mathcal{C}_f$, where $\mathcal{C}_f$ is the post-quench Chern number. However, in general $\mathcal{C}_i\neq 0$, $\mathcal{L}$ is not well-defined \cite{wilzeck,chenxin,csintegral}. Modified Chern-Simons integral can only give $(\mathcal{C}_f -\mathcal{C}_i)$ mod $2\mathcal{C}_i$ \cite{chenxin}. So new ideas are required to construct the invariant for generic quantum quenches.

We solve this problem by introducing the concept of loop unitary for quench dynamics. The unitary evolution does not have period $\pi$, $U(t=\pi)\neq\textrm{I}$. However, it can be decomposed as the product of a loop unitary $U_l$ that has time period $\pi$ and the evolution of some constant Hamiltonian \cite{roy} where we must include information about the initial state or pre-quench Hamiltonian. This motivates us to define the following loop unitary operator
\begin{eqnarray}
U_l(t)=e^{-i h t}e^{ih_0t},\label{loopU}
\end{eqnarray}
where the first term on right is the time-evolution $U$, and $h_0$ is the pre-quench Hamiltonian with $|\xi_0\rangle$ as ground state: $h_0|\xi_0\rangle=-|\xi_0\rangle$. One can check that $U_l$ indeed has period $\pi$, $U_l(0)=U_l(\pi)=\textrm{I}$. In contrast to the $\textrm{T}^3\rightarrow S^2$ mapping above, $U_l$ defines a mapping $\textrm{T}^3\rightarrow\textrm{SU(2)}$ valid for arbitrary $|\xi_0\rangle$, $h_0$ and $h$. Then, the topological invariant for quench dynamics is the 3-winding number
\begin{eqnarray}
W_3=&&\frac{1}{24\pi^2}\int_{\textrm{T}^3}d^2\bm kdt~\epsilon^{\mu\nu\rho}\times\notag\\
&&\textrm{Tr}[(U_l^{-1}\partial_{\mu}U_l)(U_l^{-1}\partial_{\nu}U_l)(U_l^{-1}\partial_{\rho}U_l)],\label{winding}
\end{eqnarray}
with $t\in[0,\pi]$. It is an integer following from homotopy group $\pi_3(\textrm{SU(2)})=\mathbb{Z}$. Note that $U_l(t)|\xi_0\rangle=e^{-it}|\xi(t)\rangle$ and $|\xi(t)\rangle$ represent the same state with different phase factors. It is easy to check $W_3$ reduces to Hopf invariant $\mathcal{L}$ when $|\xi_0\rangle$ is trivial, see \cite{SM,fhopf} for details.

{\color{blue}\textit{$\pi$-defect in phase band.}} Now we are ready to relate $W_3$ to the Chern numbers $\mathcal{C}_{f,i}$. To this end, it is convenient to diagonalize $U_l$ in its eigenbasis,
\begin{eqnarray}
U_l(t)&=&e^{i\phi(\bm k,t)}|\phi_{+}\rangle\langle\phi_{+}|+e^{-i\phi(\bm k,t)}|\phi_{-}\rangle\langle\phi_{-}|.
\end{eqnarray}
Here $\pm\phi(\bm k,t)$ (with $0\leq\phi\leq\pi)$ are called phase bands, with $|\phi_{\pm}\rangle$ the corresponding eigenstates \cite{phaseband,roy}. We define spin vector $\pm\hat{\bm m}=\langle\phi_{\mp}|\bm\sigma|\phi_{\mp}\rangle$ from the phase bands, and note that $U_l$ is homotopic \cite{SM} to a two-step evolution $U_g$ given by $U_g = e^{-i2 h t}$ for $0<t<\frac{\pi}{2}$; $U_g = -e^{i2 h_0(t-\frac{\pi}{2})}$ for $\frac{\pi}{2}<t<\pi$. Consequently $U_l$ and $U_g$ have the same topological properties and $W_3=W_3[U_g]$. This relation reveals a deep connection between quantum quench and Floquet driving described by $U_g$. Via the phase bands of $U_g$, which are easier to work with, we can prove \cite{SM} 
\begin{eqnarray}
W_3=\mathcal{C}_f-\mathcal{C}_i. \label{chern-change}
\end{eqnarray}
Hence $W_3$ gives the Chern number change and fully characterizes the quench. Compared to the Hopf invariant $\mathcal{L}$, $W_3$ is gauge-independent and valid for any initial state. Eqs. \eqref{winding} and \eqref{chern-change} are two key results of our paper.

\begin{figure}[t]
\includegraphics[width=3.35in]{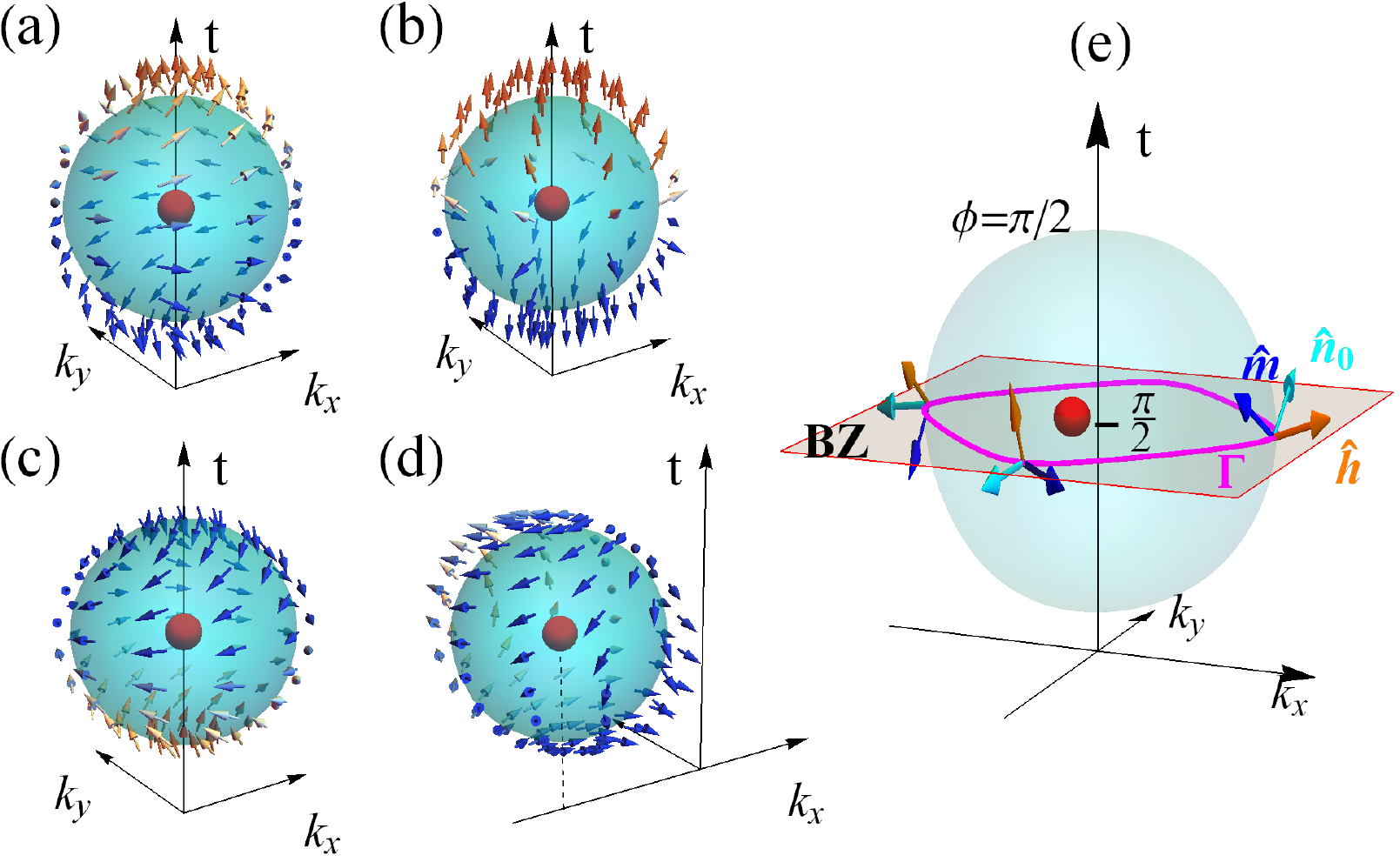}
\caption{Topological $\pi$-defects (solid red dots) as hedgehogs in the momentum-time space for quantum quenches. They are Weyl-like degeneracy points in the phase band of loop unitary $U_l$ at $\phi=\pi$. The arrow denotes the phase-band spin vector $\hat{\bm m}$ near the defect. (a)-(b): quench from a trivial initial state $|\xi_0\rangle=(0,1)^T$ to $H_{q=1}$ and $H_{q=2}$ defined in model \eqref{model}. (c)-(d): quench from the ground state of $H_{q=1}$ to $h=\sigma_z$ and $H_{q=2}$ respectively. The topological charge of the hedgehog is $+1, +2, -1, +1$ from (a) to (d), and equal to the Chern number change $\mathcal{C}_f-\mathcal{C}_i$ in each case. {$R=0$, $M=1.6$.} (e) Schematics of the DQPT curve $\Gamma$ (magenta line). It is the intersection of the $t=\frac{\pi}{2}$ plane and the equal-$\phi$ surface of $\phi=\frac{\pi}{2}$ (green) which encloses the $\pi$-defect. Along $\Gamma$, vectors $\hat{\bm n}_0$, $\hat{\bm h}$ and $\hat{\bm m}$ are perpendicular to each other.}
\label{figsky}
\end{figure}

The phase band analysis also provides an intuitive picture for the dynamical topology: $W_3$ counts the topological charge associated with the point degeneracies in the phase band \cite{phaseband,roy} at $\phi=\pi$, which we refer to as $\pi$-defects. These $\pi$-defects are located at special points in the (${\bm k}$,$t$)-space with $t=\frac{\pi}{2}$ and ${\bm k}$ determined by $\hat{\bm h}=\hat{\bm n}_0=\langle\xi_0|\bm\sigma|\xi_0\rangle$, i.e., when the post-quench Hamiltonian vector $\hat{\bm h}$ is parallel to the initial spin vector $\hat{\bm n}_0$. Expanding $U_l$ near a $\pi$-defect, we find it takes the form of Weyl Hamiltonian, e.g. $U_l(\delta k_x,\delta k_y,\delta t)=-\textrm{I}-i\delta k_i K_{ij}\sigma_j$, with $\delta k_i \in (\delta k_x,\delta k_y,\delta t)$ being deviations from the degeneracy point. Such a Weyl point carries topological charge $\sgn(\det(K))$ \cite{phaseband}. We can show that $W_3$ counts the total charges of these $\pi$-defects \cite{SM}
\begin{eqnarray}\label{charge}
W_3=\frac{1}{8\pi}\sum\nolimits_P\int_{\bm S_P} d \bm S_P\cdot~\epsilon^{ijk}(m_i\bm\nabla m_j\times\bm\nabla m_k).
\end{eqnarray}
Here $\bm S_P$ is a surface enclosing the $\pi$-defect, and the summation is over all $\pi$-defects. The phase-band spin vector $\hat{\bm m}$ forms a hedgehog [Fig. \ref{figsky}] around the $\pi$-defect.

{\color{blue}\textit{Examples.}}
To illustrate our theory, we apply it to a simple model with Hamiltonian $H_q=\bm H \bm\cdot\bm\sigma$ given by
\begin{eqnarray}\label{model}
H_x+iH_y&=&(\sin k_x+i\sin k_y)^q-R;\notag\\
H_z&=&M-\cos k_x-\cos k_y.
\end{eqnarray}
At $R=0$, the Chern number of the lower band $\mathcal{C}=q$ when $0<M<2$; $\mathcal{C}=-q$ when $-2<M<0$; and $\mathcal{C}=0$ otherwise. When $q=1$, the model reduces to the Qi-Wu-Zhang model \cite{qiwuzhang} of 2D Chern insulators recently realized using ultracold atoms in optical Raman lattices \cite{qwzexp1,qwzexp2}. We keep $R, M$ as tuning parameters and {assume $0<M<2$ below}.

Let us consider four distinct quench pathways. The first two examples start from a trivial initial state $|\xi_0\rangle=(0,1)^T$ and quench to $H_q$ above with $q=1$ and $q=2$ respectively. The $\pi$-defect for these cases is at $(k_x,k_y,t)=(0,0,\frac{\pi}{2})$, and the hedgehog patterns of the $\bm m$-vector around the defect are depicted in Figs. \ref{figsky}(a)(b). For $q=1$, expanding the loop unitary near the defect yields $U_l=-\textrm{I}+i(d\delta k_y\sigma_x-d\delta k_x\sigma_y-2\delta t\sigma_z)$ with $d={1}/({2-M})$. The charge therefore is $+1$. For $q=2$, the expansion yields a quadratic Weyl point \cite{SM} carrying charge $+2$. Next we consider quenches from the ground state of $H_q$ with $q=1$, i.e. from a topologically nontrivial initial state with $\mathcal{C}_i=1$. In the third example, the post-quench Hamiltonian is trivial $h=\sigma_z$, $\mathcal{C}_f=0$. We find a $\pi$-defect at $(0,0,\frac{\pi}{2})$ with charge $-1$, see Fig. \ref{figsky}(c). In the fourth example, the post-quench Hamiltonian is $H_q$ with $q=2$, $\mathcal{C}_f=2$. The $\pi$-defect is at $(-\arccos(M-1),0,\frac{\pi}{2})$ with charge $+1$ as depicted in Fig. \ref{figsky}(d). One can check Eq. \eqref{chern-change} indeed holds in all these four cases.

{\color{blue}\textit{DQPT curve and winding number.}} The $\pi$-defects dictating the topology of quench dynamics reside in the 3D $({\bm k}$,$t$)-space. Next we show that the dynamical singularity also manifests along a lower dimensional curve, if the pre-quench (or post-quench) Hamiltonian is trivial. This provides another intuitive picture for the dynamical topology to reveal a deep connection to DQPT. A central concept in DQPT is Loschmidt echo (LE), which measures the overlap between the initial state and time-evolved state: $\mathcal{S}(t)=|\langle\xi_0|\xi(t)\rangle|^2$ \cite{echo}. The LE is equal to the squared average of the loop unitary operator over the initial state, $\mathcal{S}(t)=|\langle\xi_0|U_l(t)|\xi_0\rangle|^2$. Geometrically, after the quench, the spin vector $\bm\xi$ of the time-evolved state precesses around $\hat{\bm h}$. By definition, DQPT occurs when LE is zero, i.e. when $\bm\xi$ becomes anti-parallel to $\hat{\bm n}_0$. This requires $t=\frac{\pi}{2}$ and $\hat{\bm h}\perp\hat{\bm n}_0$, as depicted in Fig. \ref{figsky}(e). All the points satisfying these two conditions constitute a closed curve $\Gamma$ on the $t=\frac{\pi}{2}$ plane, dubbed the DQPT curve. One can check that for points along $\Gamma$, the eigenphase of $U_l$ takes $\phi=\frac{\pi}{2}$. Hence $\Gamma$ is also the intersection of the equal-$\phi$ surface with $\phi=\frac{\pi}{2}$ and the $t=\frac{\pi}{2}$ plane, as illustrated in Fig. \ref{figsky}(e).

Suppose $\Gamma$ is parameterized by some angle $\theta\in [0,2\pi]$. Along $\Gamma$, the phase-band spin vector is given by $\hat{\bm m}=\hat{\bm h}\times\hat{\bm n}_0$, which in general tilts as $\theta$ is varied. By a unitary transformation $V^{\dag}h_0V=\sigma_z$, $U_l$ becomes $V^{\dag}U_lV=i (\tilde{m}_x\sigma_x+\tilde{m}_y\sigma_y)$. Hence the vector $\hat{\bm m}$ is fully described by polar angle $\chi=\arctan({\tilde{m}_y}/{\tilde{m}_x})$, and its round trip along $\Gamma$ is characterized by the winding number
\begin{eqnarray}
\nu=\frac{1}{2\pi}\oint_{\Gamma} d\theta~\partial_{\theta}\chi.
\label{nuw}
\end{eqnarray}
One can further prove \cite{SM} that $\nu$ coincides with $\mathcal{C}_f$, and by the chain of identities established above it is also equal to $W_3$ and $\mathcal{L}$, when the initial state is trivial.

\begin{figure}[t]
\includegraphics[width=3.35in]{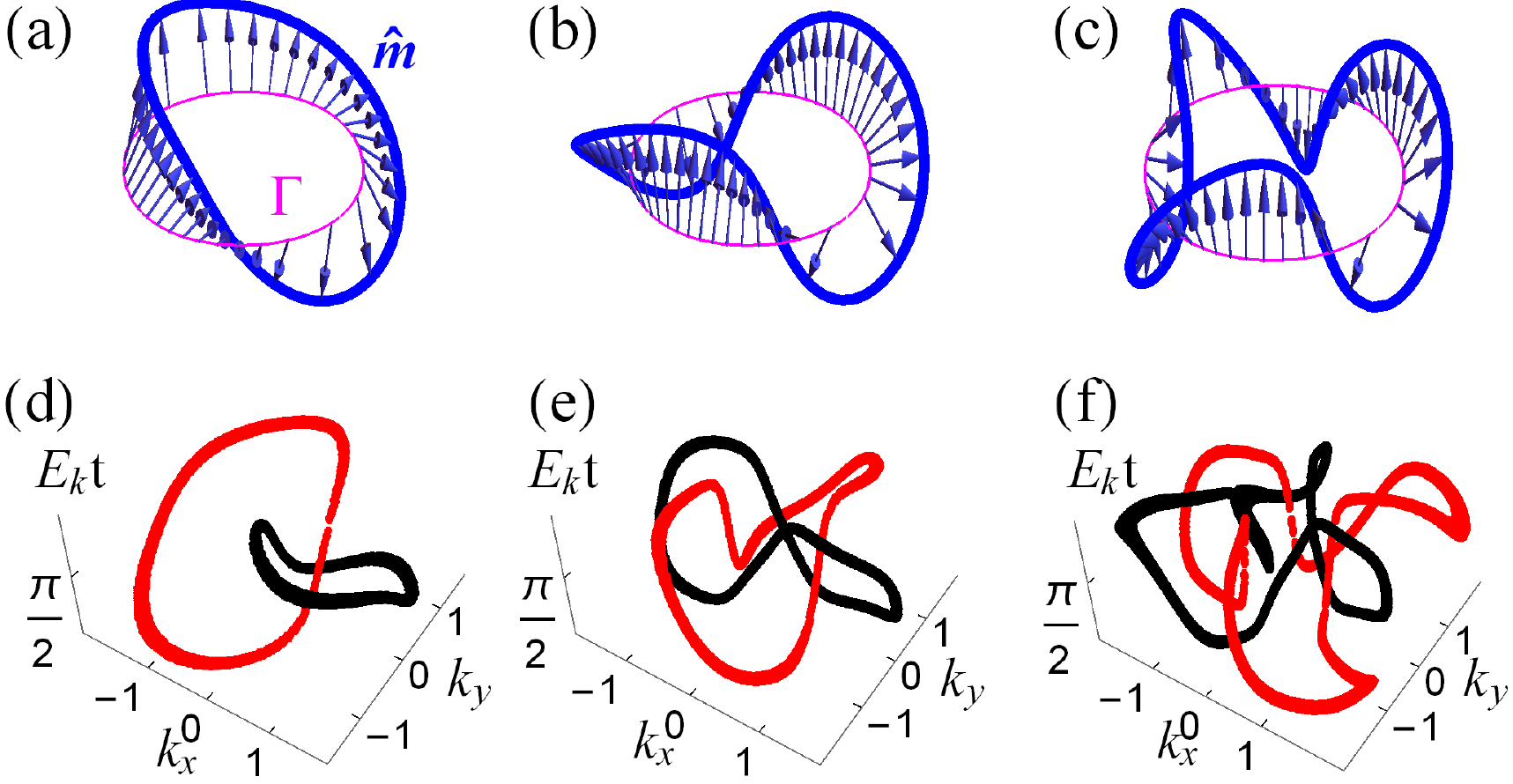}
\caption{(a)-(c): Winding of the phase-band spin vector $\hat{\bm m}$ (arrows) along the DQPT curve $\Gamma$ for quench from initial state $|\xi_0\rangle=(0,1)^T$ to $H_q$ with $q=1,2,3$ (left to right). The corresponding winding number as defined in Eq. \eqref{nuw} is $\nu=1,2,3$. (d)-(f): Hopf links in momentum-time space for the same quench to $H_q$ with $q=1,2,3$ (left to right). Shown are preimages of two time-evolved states $\pm\bm\xi=(1,0,0)$. $R=0.2$, $M=1.6$.}
\label{fig2}
\end{figure}

The winding of $\hat{\bm m}$ is illustrated in Fig. \ref{fig2} for our model \eqref{model} with initial state $|\xi_0\rangle=(0,1)^T$. The DQPT curve in this case is given by $M-\cos k_x-\cos k_y=0$. We plot $(\tilde{m}_x,\tilde{m}_y)$ as a vector on the local $x$-$y$ plane normal to $\Gamma$. Figs. \ref{fig2}(a)-(c) illustrate the winding of $\hat{\bm m}$ for $q=1,2,3$, respectively. By traveling counterclockwise along $\Gamma$, one observes that $\hat{\bm m}$ winds $1,2,3$ times within the $x$-$y$ plane, in agreement with the calculation $\nu=q$. For comparison, we have chosen two arbitrary states and plotted their preimages in the $(\bm k,t)$-space in Figs. \ref{fig2}(d)-(f). The two preimages form a Hopf link with linking number $\mathcal{L}=1,2,3$, respectively, consistent with $\nu$ above.

{\color{blue}\textit{Torus links and knots.}} Next we discuss quench from a trivial state to a Hamiltonian $H$ with Dirac points, i.e. band degeneracies at certain isolated $\bm k$ points. To this end, it is useful to imagine a torus (pipe) of unit cross-sectional radius extending along $\Gamma$. Then the end points of $\hat{\bm m}$ trace out a curve $\Gamma_+$ on the torus surface. Similarly $-\hat{\bm m}$ traces out another curve $\Gamma_-$. For fully gapped $H$, the Gauss linking number of these two closed curves, $\Gamma_\pm$, is nothing but $\nu$. If band degeneracies exist, one can verify that all Dirac points lie on $\Gamma$ projected onto the Brillouin zone. At each Dirac point, the two phase bands $|\phi_{\pm}\rangle$ become degenerate. The net effect of the band touching is the role switch $\hat{\bm m}\leftrightarrow -\hat{\bm m}$. At the Dirac points, the curves $\Gamma_\pm$ continue smoothly, only to switch their characters there, $\Gamma_+\leftrightarrow \Gamma_-$. It follows that $\Gamma_\pm$ together form either torus links or torus knots \cite{knottheory}. 

This can be illustrated by using our model Eq. \eqref{model}. When $R=[{1-(M-1)^2}]^{3/2}$ for $q=3$, one of the Weyl charges touches $\Gamma$ at the Dirac point, while the other two charges remain inside as depicted in Fig. \ref{fig3}(a). The corresponding $\Gamma_\pm$ curves are projected onto a 2D plane for clarity. In this case, the $\Gamma_+\leftrightarrow \Gamma_-$ switch happens only once. Together they form a single closed curve, a torus knot with crossing number $c_K=5$. A different scenario is shown in Fig. \ref{fig3}(b) for $q=2$ where $H$ has one pair of Dirac points. The switch occurs twice to give rise to two curves. They form a torus link with linking number $c_L=1$.

The invariants for these links and knots can be obtained as follows. Without loss of generality, we can treat $H$ with Dirac points as the critical boundary between two gapped Hamiltonians with Chern number $\mathcal{C}_<$ and $\mathcal{C}_>$ respectively as some tuning parameter (e.g. $R$) is varied. In the first scenario, {{$\mathcal{C}_>-\mathcal{C}_<=odd$}, i.e., there are odd number of Dirac points on $\Gamma$, leading to odd times of $\pm\hat{\bm m}$ switch and the formation of a torus knot [Fig. \ref{fig3}(a)]. Its crossing number \cite{knottheory} is then $c_K =2\mathcal{C}_<+\mathcal{C}_>-C_<= \mathcal{C}_< + \mathcal{C}_>$. In the second scenario, $\mathcal{C}_>-\mathcal{C}_<=even$ with even number of Dirac points on $\Gamma$. Accordingly, $\pm \hat{\bm m}$ form a torus link [Fig. \ref{fig3}(b)], and its linking number is half the crossing number \cite{knottheory}, $c_L=(\mathcal{C}_<+\mathcal{C}_>)/{2}$. These general results agree with the examples above. For $q=3$, $\mathcal{C}_<=3$ and $\mathcal{C}_>=2$ to give $c_K=5$; while for $q=2$, $\mathcal{C}_<=2$ and $\mathcal{C}_>=0$ so $c_L=1$. Thus, the concepts of loop unitary and DQPT curve also provide insights for quench to Dirac semimetals.
\begin{figure}[t]
\includegraphics[width=3.35in]{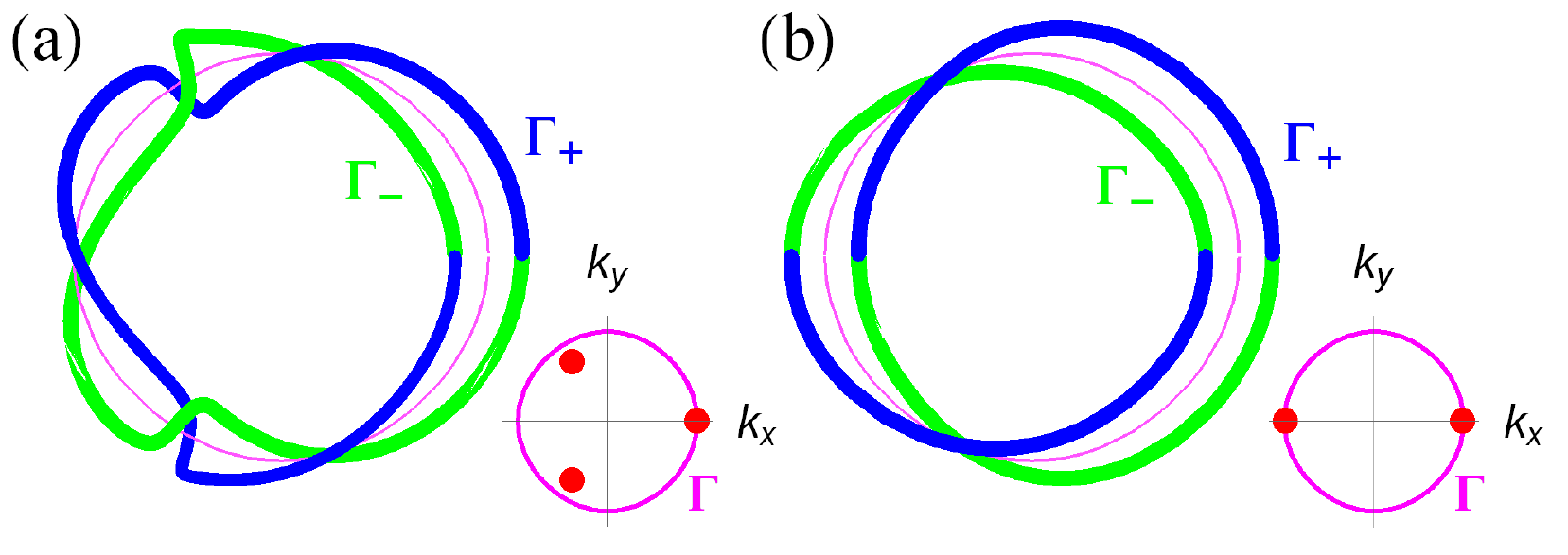}
\caption{Top view of a knot (a) and a link (b) formed by two curves $\Gamma_\pm$ traced by vector $\pm\hat{\bm m}$ along $\Gamma$ for quenches to critical $H_q$ with Dirac points. The insets show the topological charges on the $t=\pi/2$ plane. They cross $\Gamma$ exactly at the Dirac points where the two curves switch, $\Gamma_+\leftrightarrow \Gamma_-$. (a): A single curve ties into a knot with crossing number $c_K=5$, $q=3$, $R=0.512$. (b): Two closed curves form a torus link with linking number $c_L=1$, $q=2$, $R=0.64$. $M=1.6$.}
\label{fig3}
\end{figure}

{\color{blue}\textit{Outlook.}} In summary, we establish a new framework to characterize the topological properties of quench dynamics by introducing loop unitary $U_l$ and its homotopy invariant $W_3$, which goes beyond the Hopf mapping and is independent of initial state. The dynamical topology is revealed pictorially in two ways, the $\pi$-defects in the phase bands, and the winding of $\hat{\bm m}$ along the DQPT curve. The theory is generalized further to discuss the link and knot structures for quench into critical Dirac semimetal. A series of identities are proved to relate the Chern, crossing, linking, and winding number.

For concreteness, we have focused on the dynamics of two-band Hamiltonians in 2D. Our scheme based on the loop unitary however is general. For example, we have applied it to obtain the $\mathbb{Z}_2$ dynamical invariant for the quench dynamics of Hopf insulators \cite{hhphopf} in 3D, and characterize the quench dynamics of the four-band Bernevig-Hughes-Zhang model \cite{SM}. For these more complicated cases, the loop unitary must be chosen to be time-periodic to ensure a closed base manifold, and the additional symmetry constraints must be properly taken into account \cite{SM} in constructing the desired dynamical topological invariants. Through the concept of loop unitary and homotopy relation $U_l\sim U_g$, our work also reveals an intrinsic connection between quench and Floquet dynamics \cite{roy,wangzhong}. It suggests that a wealth of link and knot structures will also emerge in $({\bm k},t)$-space in Floquet systems. For example, Hopf link was recently shown to appear in periodically driven 2D systems \cite{fhopf}.

Recent experiments have begun to quantitatively access quench dynamics via time- and momentum-resolved tomography \cite{azi1,azi2,tomograph1,tomograph2}. In particular, spatiotemporal Hopf links \cite{quenchexp2,quenchexp3,quenchexp4} after quantum quench have been observed. The $\pi$-defect in the phase band can be observed via the same Bloch-state tomography technique. For example, its topological charge, which is quantized and protected against small perturbations, can be extracted locally from the nearby tomography points \cite{chargeexp}. The location of the DQPT curve and the winding along it can be verified by tracing the emergent dynamical vortices in momentum space as done in Ref. \cite{quenchexp1}. For more details, see Supplementary Materials \cite{SM}.

\begin{acknowledgments}
This work is supported by AFOSR Grant No. FA9550-16-1-0006 and NSF Grant No. PHY-1707484.
\end{acknowledgments}

\clearpage
\onecolumngrid
\appendix
\section{Supplementary Materials}
In this Supplementary material, we provide details on (I) the proof of the relation between $W_3$ and Hopf invariant $\mathcal{L}$ with trivial initial state; (II) the homotopy relation between loop unitary $U_l$ and $U_g$; (III) phase-band representation of $W_3$; (IV) $\pi$-defect and Weyl charge; (V) the winding topology of DQPT curve $\Gamma$; (VI) dynamical topology of the Bernevig-Hughes-Zhang model; (VII) and experimental detection of DQPT curve and topological charge.

\subsection{I. $W_3$ and Hopf invariant $\mathcal{L}$}
In the main text, we have mentioned that for a trivial initial state $|\xi_0\rangle$, $W_3=\mathcal{L}$. To prove this, we utilize the property of the loop unitary: $U_l|\xi_0\rangle=e^{-it}|\xi(t)\rangle$ represents the same quantum state with $|\xi(t)\rangle$ on the Bloch sphere. Then the proof follows that of Ref. \cite{fhopf}. Let us expand the 3-winding number $W_3$ by inserting a complete set of $\bm k$-independent basis $|\psi_a\rangle$ ($a=1,2$, and $|\psi_1\rangle\langle\psi_1|+|\psi_2\rangle\langle\psi_2|=\textrm{I}$) as
\begin{eqnarray}\label{w=hopf}
W_3&=&\frac{1}{24\pi^2}\int_{\textrm{T}^3}d^2\bm kdt~\epsilon^{\mu\nu\rho}~\textrm{Tr}[(U_l^{-1}\partial_{\mu}U_l)(U_l^{-1}\partial_{\nu}U_l))(U_l^{-1}\partial_{\rho}U_l)]\notag\\
&=&-\frac{1}{24\pi^2}\int_{\textrm{T}^3}d^2\bm kdt~\epsilon^{\mu\nu\rho}~\textrm{Tr}[U_l^{\dag}\partial_{\mu}U_l\partial_{\nu}U_l^{\dag}\partial_{\rho} U_l]\notag\\
&=&-\frac{1}{24\pi^2}\int_{\textrm{T}^3}d^2\bm kdt~\epsilon^{\mu\nu\rho}\sum_{a,b,c,d}\langle\psi_a|U_l^{\dag}|\psi_b\rangle\langle\psi_b|\partial_{\mu}U_l|\psi_c\rangle\langle\psi_c|\partial_{\nu}U_l^{\dag}|\psi_d\rangle\langle\psi_d|\partial_{\rho} U_l|\psi_a\rangle\notag\\
&=&-\frac{1}{24\pi^2}\int_{\textrm{T}^3}d^2\bm kdt~\epsilon^{\mu\nu\rho}\sum_{a,c}\langle\Psi_a|\partial_{\mu}\Psi_c\rangle\langle\partial_{\nu}\Psi_c|\partial_{\rho}\Psi_a\rangle.
\end{eqnarray}
In the above equations, $a,b,c,d=1,2$ and $|\Psi_a\rangle\equiv U_l|\psi_a\rangle$. In its most general form, the two-component spinors take $|\Psi_1\rangle=(z_1,z_2)^T$, with $|z_1|^2+|z_2|^2=1$, and $|\Psi_2\rangle=(-z_2^*,z_1^*)^T$. The last line of Eq. (\ref{w=hopf}) contains four terms. It is easy to check the two terms with $a=c=1$ and $a=c=2$ are equal:
\begin{eqnarray}
W_3^{(1,1)}&=&\epsilon^{\mu\nu\rho}\langle\Psi_1|\partial_{\mu}\Psi_1\rangle\langle\partial_{\nu}\Psi_1|\partial_{\rho}\Psi_1\rangle\notag\\
&=&\epsilon^{\mu\nu\rho}(z_1^*,z_2^*)(\partial_{\mu} z_1,\partial_{\mu} z_2)^T(\partial_{\nu} z_1^*,\partial_{\nu} z_2^*)(\partial_{\rho} z_1,\partial_{\rho} z_2)^T\notag\\
&=&\epsilon^{\mu\nu\rho}(z_1^*\partial_{\mu} z_1+z_2^*\partial_{\mu} z_2)(\partial_{\nu} z_1^*\partial_{\rho} z_1+\partial_{\nu} z_2^*\partial_{\rho} z_2)\notag\\
&=&\epsilon^{\mu\nu\rho}(z_1^*\partial_{\mu} z_1\partial_{\nu} z_2^*\partial_{\rho} z_2+z_2^*\partial_{\mu} z_2\partial_{\nu} z_1^*\partial_{\rho} z_1);\\
W_3^{(2,2)}&=&\epsilon^{\mu\nu\rho}\langle\Psi_2|\partial_{\mu}\Psi_2\rangle\langle\partial_{\nu}\Psi_2|\partial_{\rho}\Psi_2\rangle\notag\\
&=&\epsilon^{\mu\nu\rho}(-z_2,z_1)(-\partial_{\mu} z_2^*,\partial_{\mu} z_1^*)^T(-\partial_{\nu} z_2,\partial_{\nu} z_1)(-\partial_{\rho} z_2^*,\partial_{\rho} z_1^*)^T\notag\\
&=&\epsilon^{\mu\nu\rho}(z_2\partial_{\mu} z_2^*+z_1\partial_{\mu} z_1^*)(\partial_{\nu} z_2\partial_{\rho} z_2^*+\partial_{\nu} z_1\partial_{\rho} z_1^*)\notag\\
&=&\epsilon^{\mu\nu\rho}(z_2\partial_{\mu} z_2^*\partial_{\nu} z_1\partial_{\rho} z_1^*+z_1\partial_{\mu} z_1^*\partial_{\nu} z_2\partial_{\rho} z_2^*)\notag\\
&=&\epsilon^{\mu\nu\rho}(-z_2^*\partial_{\mu} z_2\partial_{\nu} z_1\partial_{\rho} z_1^*-z_1^*\partial_{\mu} z_1\partial_{\nu} z_2\partial_{\rho} z_2^*)=W_3^{(1,1)}.
\end{eqnarray}
In the above calculations, the symmetric terms of $(\mu\nu\rho)$, e.g., $z_1^*\partial_{\nu} z_1^*\partial_{\mu} z_1\partial_{\rho} z_1$ vanish due to the antisymmetric Levi-Civita symbol. The mixed terms with $a\neq c$ can be represented using the above unmixed terms as
\begin{eqnarray}
W_3^{(1,2)}&=&\epsilon^{\mu\nu\rho}\langle\Psi_1|\partial_{\mu}\Psi_2\rangle\langle\partial_{\nu}\Psi_2|\partial_{\rho}\Psi_1\rangle\notag\\
&=&\epsilon^{\mu\nu\rho}(z_1^*,z_2^*)(-\partial_{\mu} z_2^*,\partial_{\mu} z_1^*)^T(-\partial_{\nu} z_2,\partial_{\nu} z_1)(\partial_{\rho} z_1,\partial_{\rho} z_2)^T\notag\\
&=&\epsilon^{\mu\nu\rho}(-z_1^*\partial_{\mu} z_2^*+z_2^*\partial_{\mu} z_1^*)(-\partial_{\nu} z_2\partial_{\rho} z_1+\partial_{\nu} z_1\partial_{\rho} z_2)\notag\\
&=&\epsilon^{\mu\nu\rho}(z_1^*\partial_{\mu} z_2^*\partial_{\nu} z_2\partial_{\rho} z_1-z_2^*\partial_{\mu} z_1^*\partial_{\nu} z_2\partial_{\rho} z_1-z_1^*\partial_{\mu} z_2^*\partial_{\nu} z_1\partial_{\rho} z_2+z_2^*\partial_{\mu} z_1^*\partial_{\nu} z_1\partial_{\rho} z_2)\notag\\&=&2W_3^{(1,1)}.
\end{eqnarray}
Similarly $W_3^{(2,1)}=2W_3^{(1,1)}$, yielding $W_3^{(1,1)}+W_3^{(2,2)}+W_3^{(1,2)}+W_3^{(2,1)}=6W_3^{(1,1)}$. $W_3$ can then be rewritten as
\begin{eqnarray}
W_3=-\frac{1}{4\pi^2}\int_{\textrm{T}^3}d^2\bm kdt~\epsilon^{\mu\nu\rho}\langle\Psi_1|\partial_{\mu}\Psi_1\rangle\langle\partial_{\nu}\Psi_1|\partial_{\rho}\Psi_1\rangle.
\end{eqnarray}
Now let us compare $W_3$ with the Hopf invariant
\begin{eqnarray}
\mathcal{L}=\frac{1}{4\pi^2}\int_{\textrm{T}^3}d^2\bm kdt~\epsilon^{\mu\nu\rho}\mathcal{A}_{\mu}\partial_{\nu}\mathcal{A}_{\rho}=-\frac{1}{4\pi^2}\int_{\textrm{T}^3}d^2\bm kdt~\epsilon^{\mu\nu\rho}\langle\xi|\partial_{\mu}\xi\rangle\langle\partial_{\nu}\xi|\partial_{\rho}\xi\rangle.
\end{eqnarray}
The physical meaning of $\mathcal{L}$ is the linking number of preimages of any two quantum states on the Bloch sphere. For quantum quench from trivial initial state $|\xi_0\rangle$, we can continuously deform $|\xi_0\rangle$ to be $\bm k$-independent. By choosing $|\psi_1\rangle=|\xi_0\rangle$, we have $|\Psi_1\rangle=U_l|\psi_1\rangle\sim|\xi(t)\rangle$. Further note that $W_3$ is a gauge-independent invariant, regardless of the phase factor of $|\Psi_1\rangle$, hence we arrive at
\begin{eqnarray}
W_3=\mathcal{L}.
\end{eqnarray}
The relation between the two homotopy invariant $W_3$ and $\mathcal{L}$ can be further understood from a mathematical point of view. As $U_l|\xi_0\rangle\sim|\xi(t)\rangle$, the loop unitary defines a fiber bundle on $S^2$. $W_3=\mathcal{L}$ is ensured by the homotopy relation $\pi_3(\textrm{SU(2)})=\pi_3(S^3)=\pi_3(S^2)=\mathbb{Z}$.

\subsection{II. Proof of homotopy relation of loop unitary}
We first explain the mathematical meaning of homotopy between unitary matrices: two unitary matrices $U_M(t)$ and $U_N(t)$ are homotopic when there exists a continuous matrix function $U(t,\lambda)$ with $\lambda\in[0,1]$, such that $U(t,\lambda=0)=U_M(t)$, $U(\lambda=1)=U_N(t)$ and $U(t,\lambda)$ preserves the (relevant) band-gap (here we focus on the $\pi$-gap) for all $\lambda$. Our proof is based on the fact that any unitary evolution $U(t)$ can be decomposed into two parts, a loop unitary part and a constant evolution part \cite{roy}, denoted as $U(t)=L(t)*C(t)$. Here $L(t)$ is the loop unitary $L(t=0)=L(t=\pi)=\textrm{I}$, and $C(t)$ is the constant evolution. $*$ indicates the standard composition of path \cite{nakahara,roy}. The above decomposition is unique up to homotopy. Consider the two-step time evolution parameterized by $\lambda_1\in[0,1]$:
\begin{eqnarray}\label{path1}
U(t,\lambda_1) =\Big\{\begin{array}{l l} e^{-2 i h t}e^{2 i \lambda_1 h_0 t}, ~~~~~~\quad 0<t<\frac{\pi}{2}; \\ -e^{i \pi \lambda_1 h_0}e^{i h_0(t-\frac{\pi}{2})}, \quad \frac{\pi}{2}<t<\pi.\\ \end{array}
\end{eqnarray}
Let us start with $U(t,\lambda_1=1)$. At the first half period ($0<t<\frac{\pi}{2}$), it is the desired loop unitary $U_l$ defined in the main text. The second half period is a constant evolution. Hence Eq. (\ref{path1}) by definition is the decomposition of $U(t,\lambda_1=1)$, with $U_l$ the corresponding loop unitary. Note that $U(t=\pi,\lambda_1)=e^{i\pi[(\frac{1}{2}+\lambda_1)h_0+1]}$, with its spectrum (phase band) given by $e^{i\pi[\pm(\frac{1}{2}+\lambda_1)+1]}$. We can clearly see the $\pi$-gap is preserved in the whole path $\lambda_1\in[0,1]$. Hence $U(t,\lambda_1=0)$ and $U(t,\lambda_1=1)$ are homotopic. According to the uniqueness of decomposition \cite{roy}, $U_l$ is homotopic to the loop unitary generated by $U(t,\lambda_1=0)$. We write down $U(t,\lambda_1=0)$ as follows:
\begin{eqnarray}
U(t,\lambda_1=0) =\Big\{\begin{array}{l l} e^{-2 i h t}, ~~~~~~\quad 0<t<\frac{\pi}{2}; \\ -e^{i h_0(t-\frac{\pi}{2})}, \quad \frac{\pi}{2}<t<\pi.\\ \end{array}
\end{eqnarray}
Now we construct the second path parameterized by $\lambda_2\in[0,1]$:
\begin{eqnarray}\label{path2}
U(t,\lambda_1=0,\lambda_2) =\Big\{\begin{array}{l l} e^{-2 i h t},~~~~~~~~~~~~~~\quad 0<t<\frac{\pi}{2}; \\ -e^{i (\lambda_2+1)h_0(t-\frac{\pi}{2})}, \quad \frac{\pi}{2}<t<\pi.\\ \end{array}
\end{eqnarray}
It is easy to check $U(t,\lambda_1=0,\lambda_2=0)=U(t,\lambda_1=0)$ and $U(t,\lambda_1=0,\lambda_2=1)=U_g$, which is the required two-step evolution in the main text. The $\pi$-gap is preserved for all $\lambda_2\in[0,1]$. Finally, we conclude $U_l$ is homotopic to $U_g$. We note that in the proof above, the loop unitary $U_l$ must be time-periodic. The homotopy relation can be extended to multi-band cases and other symmetry classes, as long as the relevant $\pi$-gap is preserved during the whole parameter path.

\subsection{III. Phase-band representation of $W_3$}
This part is devoted to the derivation of Eqs. (5)(6) in the main text, which are the phase-band representations of the 3-winding number $W_3$. In its eigenbasis, the loop unitary takes
\begin{eqnarray}
U_l&=&e^{i\phi}|\phi_{+}\rangle\langle\phi_{+}|+e^{-i\phi}|\phi_{-}\rangle\langle\phi_{-}|\notag\\
&=&\cos\phi-i\sin \phi~\hat{\bm m}\cdot\bm\sigma\equiv n_0\sigma_0-in_j\sigma_j,
\end{eqnarray}
with $\hat{\bm m}=\langle\phi_-|\bm\sigma|\phi_-\rangle$. $\sum_{j=0}^3(n_j)^2=1$. As $\bm n=(n_0,n_1,n_2,n_3)$ can be represented by a point on $S^3$, the loop unitary also provides a mapping from $\textrm{T}^3$ to $S^3$. Note $U^{\dag}_l=n_0\sigma_0+in_j\sigma_j$, $W_3$ can be recast into
\begin{eqnarray}
W_3&=&\frac{1}{24\pi^2}\int_{\textrm{T}^3}d^2\bm kdt~\epsilon^{\mu\nu\rho}~\textrm{Tr}[(U_l^{-1}\partial_{\mu}U_l)(U_l^{-1}\partial_{\nu}U_l))(U_l^{-1}\partial_{\rho}U_l)]\notag\\
&=&-\frac{1}{24\pi^2}\int_{\textrm{T}^3}d^2\bm kdt~\epsilon^{\mu\nu\rho}~\textrm{Tr}[U_l^{\dag}\partial_{\mu}U_l\partial_{\nu}U_l^{\dag}\partial_{\rho} U_l]\notag\\
&=&-\frac{1}{24\pi^2}\int_{\textrm{T}^3}d^2\bm kdt~\epsilon^{\mu\nu\rho}~(-1)^{2-\delta_{b,0}-\delta_{d,0}}(i)^{F}n_a\partial_{\mu} n_b\partial_{\nu} n_c\partial_{\rho} n_d\textrm{Tr}[\sigma_a\sigma_b\sigma_c\sigma_d],
\end{eqnarray}
where $(abcd)$ take values in $(0123)$. $F=4-\delta_{a,0}-\delta_{b,0}-\delta_{c,0}-\delta_{d,0}$ counts the number of nonzero elements in the set $(abcd)$. $\delta_{a,0}=1$ if $a=0$ and $\delta_{a,0}=0$ otherwise. There are many terms in the above equation. However, the contributions to $W_3$ come solely from the terms with $F=3$, i.e., the set $(abcd)$ contains 1 zero. (1) The 4-zero terms ($F=0$) and 3-zero terms ($F=1$) vanish due to the anti-symmetric Levi-Civita symbol; (2) For the 2-zero terms ($F=2$), the trace requires the other two nonzero elements to be the same; (3) For $F=4$, the trace requires two of them (and the other two) to be equal. For cases (2)(3), the contributions to $W_3$ vanish due to the anti-symmetric Levi-Civita symbol. Now let us calculate the trace for the 1-zero terms with $F=3$. The Pauli matrices satisfy
\begin{eqnarray}
[\sigma_{\mu},\sigma_{\nu}]&=&2i\epsilon_{0\mu\nu\rho}\sigma_{\rho},\\
\{\sigma_{\mu},\sigma_{\nu}\}&=&2(\delta_{\mu\nu}\sigma_0+\delta_{0\mu}\sigma_{\nu}+\delta_{\nu 0}\sigma_{\mu})-4\delta_{\mu 0}\delta_{\nu 0}\sigma_0\equiv 2\delta_{|\mu\nu}\sigma_{0|}-4\delta_{\mu 0}\delta_{\nu 0}\sigma_0.
\end{eqnarray}
Above we have defined the cyclic summations $||$. Hence we have
\begin{eqnarray}
\sigma_{\mu}\sigma_{\nu}=i\epsilon_{0\mu\nu\rho}\sigma_{\rho}+\delta_{|\mu\nu}\sigma_{0|}-2\delta_{\mu 0}\delta_{\nu 0}\sigma_0,
\end{eqnarray}
For the 1-zero term with $d=0$,
\begin{eqnarray}
\textrm{Tr}[\sigma_a\sigma_b\sigma_c]=\textrm{Tr}[(i\epsilon_{0abf}\sigma_{f}\sigma_c+\delta_{|ab}\sigma_{0|}\sigma_c-2\delta_{a 0}\delta_{b 0}\sigma_0\sigma_c]=2i\epsilon_{0abc}.
\end{eqnarray}
Similar analysis applies to the other three 1-zero terms. By further taking into account the sign and $(i)^F$ factor, we can get
\begin{eqnarray}\label{windingv}
W_3=-\frac{1}{12\pi^2}\int_{\textrm{T}^3}d^2\bm kdt~~\epsilon_{abcd}\epsilon^{\mu\nu\rho}n_a\partial_{\mu} n_b\partial_{\nu} n_c\partial_{\rho} n_d.
\end{eqnarray}
One can see that the integrand is the unit surface element of $S^3$. Hence $W_3$ can also be interpreted as the winding number of homotopy group $\pi_3(S^3)$. We expand the four different terms in Eq. (\ref{windingv}) (note a factor of $6$).
\begin{eqnarray}
W_3=-\frac{1}{2\pi^2}\int_{\textrm{T}^3}d^2\bm kdt~~\large[n_0\bm\nabla n_1\cdot(\bm\nabla n_2\times \bm\nabla n_3)-\bm\nabla n_0\cdot (n_1\bm\nabla n_2\times \bm\nabla n_3-n_2\bm\nabla n_1\times \bm\nabla n_3+n_3\bm\nabla n_1\times \bm\nabla n_2)\large].
\end{eqnarray}
Each integrand can be calculated as
\begin{eqnarray}
n_0\bm\nabla n_1\cdot(\bm\nabla n_2\times \bm\nabla n_3)&=&\cos\phi\sin^3\phi\bm\nabla m_1\cdot(\bm\nabla m_2\times\bm\nabla m_3)\notag\\
&&+\sin^2\phi\cos^2\phi\bm\nabla\phi\cdot(m_1\bm\nabla m_2\times\bm\nabla m_3-m_2\bm\nabla m_1\times\bm\nabla m_3+m_3\bm\nabla m_1\times\bm\nabla m_2);\notag\\
n_i\bm\nabla n_0\cdot(\bm\nabla n_j\times \bm\nabla n_k)&=&-\sin^4\phi\bm\nabla\phi\cdot(m_i\bm\nabla m_j\times\bm\nabla m_k).
\end{eqnarray}
Now we argue the contributions to $W_3$ from the first line vanish. This is because $W_3$ is a topological invariant and remains invariant by continuously deforming the phase bands as long as the topological charges ($\pi$-defects) of the loop unitary $U_l$ are kept. A simple choice is to deform all the phase bands to be flat bands \cite{qq8,phaseband} with $\phi=0$ except for small regions near the defects (The phase band approaches to a step-like function.) This band-flattening procedure will greatly simplify the calculations. It is clear the terms without $\phi$-derivative $\bm\nabla\phi$ vanish after this deformation. Combining the terms in the second and third lines together, we have
\begin{eqnarray}
W_3=-\frac{1}{4\pi^2}\int_{\textrm{T}^3}d^2\bm kdt~\epsilon^{ijk}\sin^2\phi\bm\nabla\phi\cdot(m_i\bm\nabla m_j\times\bm\nabla m_k),
\end{eqnarray}
As the loop unitary $U_l$ is homotopic to the two-step evolution $U_g$, they have the same invariant $W_3=W_3[U_g]$. The latter can be easily calculated due to its simple form. The phase band of $U_g$ is: for $0<t<\frac{\pi}{2}$, $\phi=2t$; and for $\frac{\pi}{2}<t<\pi$, $\phi=\pi-2(t-\frac{\pi}{2})$. And in each time interval, the phase-band spin vector of $U_g$ is given by the corresponding eigenstate of the post- or pre-quench Hamiltonian. Explicitly,
\begin{eqnarray}
W_3[U_g]&=&-\frac{1}{4\pi^2}\Big[\int_0^{\frac{\pi}{2}}dt\int_{\textrm{BZ}}~d^2\bm k\cdot~\epsilon^{ijk}\sin^2\phi(m_i\bm\nabla m_j\times\bm\nabla m_k)+\int_{\frac{\pi}{2}}^{\pi}dt\int_{\textrm{BZ}}~d^2\bm k\cdot~\epsilon^{ijk}\sin^2\phi(m_i\bm\nabla m_j\times\bm\nabla m_k)\Big]\notag\\
&=&-\frac{1}{4\pi^2}\Big[\int_0^{\pi}d\phi\sin^2\phi\int_{\textrm{BZ}}~d^2\bm k\cdot~\epsilon^{ijk}(m_i\bm\nabla m_j\times\bm\nabla m_k)+\int_{\pi}^{0}d\phi\sin^2\phi\int_{\textrm{BZ}}~d^2\bm k\cdot~\epsilon^{ijk}(m_i\bm\nabla m_j\times\bm\nabla m_k)\Big]\notag\\
&=&-\frac{1}{8\pi}\int_{\textrm{BZ}}~d^2\bm k\cdot~\epsilon^{ijk}(m_i\bm\nabla m_j\times\bm\nabla m_k)|_{t<\frac{\pi}{2}}+\frac{1}{8\pi}\int_{\textrm{BZ}}~d^2\bm k\cdot~\epsilon^{ijk}(m_i\bm\nabla m_j\times\bm\nabla m_k)|_{t>\frac{\pi}{2}}\notag\\
&=&-(-\mathcal{C}_f)-\mathcal{C}_i=\mathcal{C}_f-\mathcal{C}_i.
\end{eqnarray}
The phase band provides a simple and intuitive understanding of the dynamical topology. To show this, we transform the $(\bm k,t)$-coordinate in the integral to coordinate $(x_1,x_2,\phi)$, with $(x_1,x_2)$ parameterizes the 2D manifold with constant $\phi$. Each equal-$\phi$ manifold is 2D closed surface (may have multiple components). In general, this coordinate transformation is not (one to one) on the entire $\textrm{T}^3$ space, and we have to divide the equal-$\phi$ surface into a union of patches $P$, each with their own local coordinates $(x_{1P},x_{2P},\phi)$ and integration surface $\bm S_P$. Using this coordinate transformation, $W_3$ can be rewritten as
\begin{eqnarray}
W_3&=&-\frac{1}{4\pi^2}\int_{\textrm{T}^3}d^2\bm kdt~\epsilon^{ijk}\sin^2\phi\bm\nabla\phi\cdot(m_i\bm\nabla m_j\times\bm\nabla m_k)\notag\\
&=&\frac{1}{4\pi^2}\sum_P \int_0^{\pi}d\phi~\sin^2\phi\int_{\bm S_P} d \bm S_P\cdot~\epsilon^{ijk}(m_i\bm\nabla m_j\times\bm\nabla m_k)\notag\\
&=&\frac{1}{8\pi}\sum_P\int_{\bm S_P} d \bm S_P\cdot~\epsilon^{ijk}(m_i\bm\nabla m_j\times\bm\nabla m_k).
\end{eqnarray}
In the above equations, the surface element $d\bm S_P$ is pointing along the direction with decreasing $\phi$ (i.e., outwards from the $\pi$-defect). The last line is exactly the wrapping number (Chern number) of the phase-band spin vector $\hat{\bm m}$ over the surface enclosing the $\pi$-defect.

\subsection{IV. $\pi$-defect and Weyl charge}
For the quantum quench from initial state $|\xi_0\rangle=(0,1)^T$ with post-quench Hamiltonian $H_{q=2}$ (model (7) in the main text), the $\pi$-defect is located at $(k_x,k_y,t)=(0,0,\frac{\pi}{2})$. The DQPT curve is $\Gamma:~M-\cos k_x-\cos k_y=0$, as depicted in Fig. \ref{dqpt_sm}(a). At the defect, $H_{q=2}=(M-2)\sigma_z$. We expand the loop unitary near this defect ($d=\frac{1}{2-M}$),
\begin{eqnarray}
U_l(\delta k_x,\delta k_y,\delta t)&=&e^{-i[d(\delta k_x^2-\delta k_y^2)\sigma_x+2d\delta k_x\delta k_y\sigma_y+\frac{d}{2}(\delta k_x^2+\delta k_y^2)\sigma_z-\sigma_z](\frac{\pi}{2}+\delta t)}e^{i\sigma_z(\frac{\pi}{2}+\delta t)}\notag\\
&=&[-\delta t-id(\delta k_x^2-\delta k_y^2)\sigma_x-2id\delta k_x\delta k_y\sigma_y-i\frac{d}{2}(\delta k_x^2+\delta k_y^2)\sigma_z+i\sigma_z](-\delta t+i\sigma_z)\notag\\
&=&-\textrm{I}+O(\delta k^2)-i[-2d\delta k_x\delta k_y\sigma_x+d(\delta k_x^2-\delta k_y^2)\sigma_y+2\delta t\sigma_z].
\end{eqnarray}
It is clear the $\pi$-defect is a quadratic Weyl point with charge $+2$.
\begin{figure}[h]
\includegraphics[width=3in]{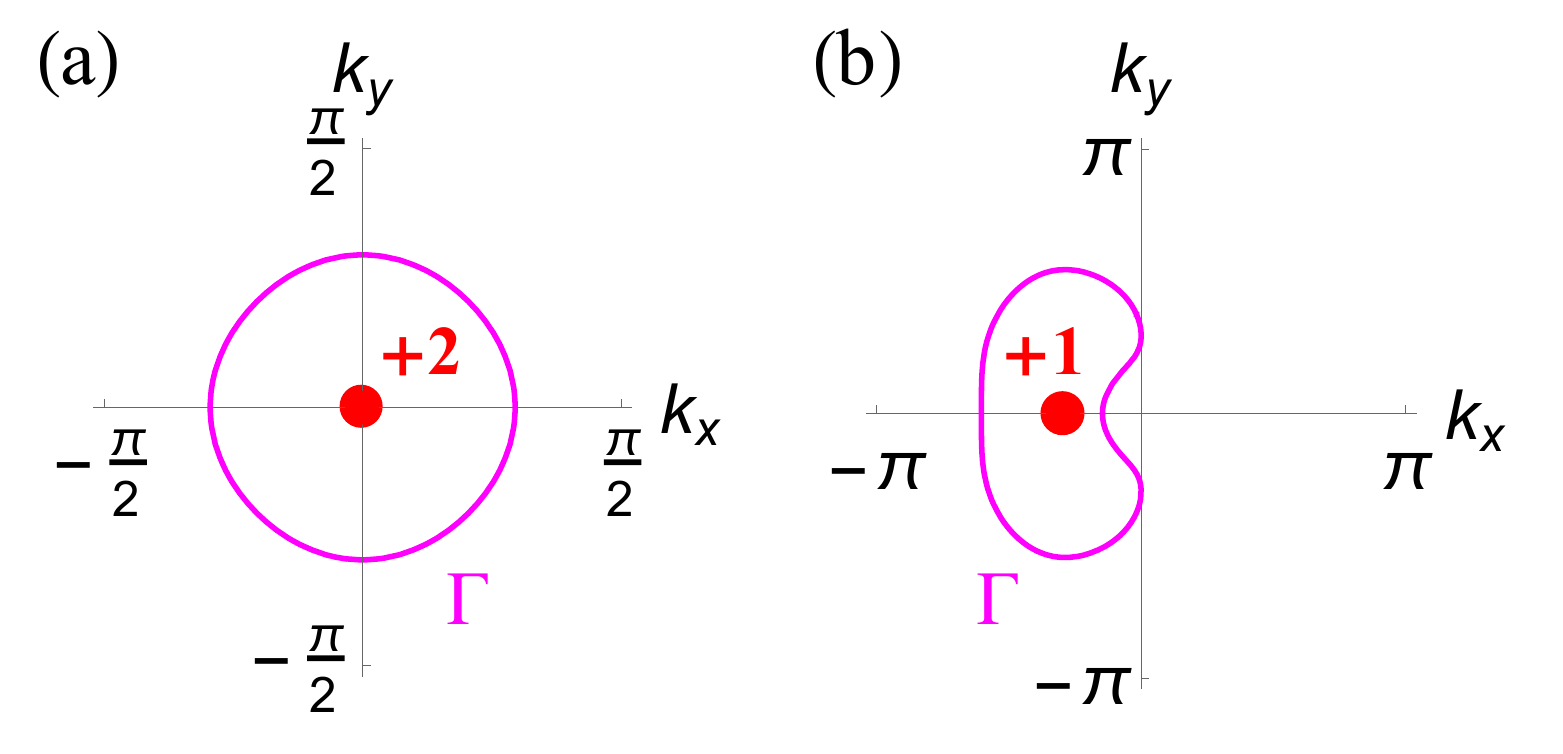}
\caption{$\pi$-defect and DQPT curve $\Gamma$. (a) Quench from $\mathcal{C}_i=0$ to $\mathcal{C}_f=2$. (b) Quench from $\mathcal{C}_i=1$ to $\mathcal{C}_f=2$. The red dots represent the $\pi$-defect and the magenta curve is DQPT curve $\Gamma$. $M=1.6$.}
\label{dqpt_sm}
\end{figure}

Now let us turn to another quantum quench, starting from the ground state of $H_{q=1}$ to $H_{q=2}$. As depicted in Fig. \ref{dqpt_sm}(b), the $\pi$-defect is located at $(k_x,k_y,t)=(-\arccos(M-1),0,\frac{\pi}{2})$. The DQPT curve is given by
\begin{eqnarray}
\Gamma:~\sin^3 k_x+\sin k_x\sin^2 k_y+(M-\cos k_x-\cos k_y)^2=0.
\end{eqnarray}
Denote $x_0=-\arccos(M-1)$. The expansion near the defect takes $h(\delta k_x,\delta k_y,\delta t)=\sigma_x+\frac{1}{\sin x_0}(2\cos x_0\delta k_x\sigma_x+2\delta k_y\sigma_y+\delta k_x\sigma_z)$, $h_0(\delta k_x,\delta k_y,\delta t)=-\sigma_x-\frac{1}{\sin x_0}(\cos x_0\delta k_x\sigma_x+\delta k_y\sigma_y+\sin x_0\delta k_x\sigma_z)$,
\begin{eqnarray}
U_l(\delta k_x,\delta k_y,\delta t)&=&e^{-ih(\delta k_x,\delta k_y,\delta t)(\frac{\pi}{2}+\delta t)}e^{ih_0(\delta k_x,\delta k_y,\delta t)(\frac{\pi}{2}+\delta t)}\notag\\
&=&2i\delta t\sigma_x+h(\delta k_x,\delta k_y,\delta t)h_0(\delta k_x,\delta k_y,\delta t)\notag\\
&=&-\textrm{I}+i(2\delta t\sigma_x-\frac{1-\sin x_0}{\sin x_0}\delta k_x\sigma_y+\frac{1}{\sin x_0}\delta k_y\sigma_z).
\end{eqnarray}
It is clear the $\pi$-defect is a linear Weyl point with charge $+1$.

\subsection{V. Winding topology of DQPT curve} In this part, we show the winding number $\nu$ (Eq. (8) in the main text) coincides with $\mathcal{C}_f$ when the initial state is trivial, $\mathcal{C}_i=0$. To this end, we consider Hamiltonian $\tilde{h}\equiv h_0hh_0=2(\hat{\bm h}_0\cdot\hat{\bm h})h_0-h$. It is obvious $\tilde{h}$ and $h$ have the same topological properties. In the chiral basis set by $h_0$ (under the unitary transformation $V$ which diagonalizes $h_0$), $V^{\dag}\tilde{h}V=2(\hat{\bm h}_0\cdot\hat{\bm h})\sigma_z-V^{\dag}hV$. Along $\Gamma$, $\hat{\bm h}_0\cdot\hat{\bm h}=0$, $V^{\dag}hV$ has only $\sigma_x, \sigma_y$ components. Hence $\Gamma$ is the band inversion surface \cite{qq8}, with the Chern number $\mathcal{C}_f$ equal to the winding number of $V^{\dag}hV$ along $\Gamma$. Further note $V^{\dag}U_lV=V^{\dag}hh_0V=V^{\dag}hVV^{\dag}h_0V=V^{\dag}hV\sigma_z$ along $\Gamma$. Hence the winding number defined in the main text is $\nu=\mathcal{C}_f$. The same argument applies to the case with trivial post-quench Hamiltonian, where $V$ is chosen as the unitary transformation $V^{\dag}hV=\sigma_z$.

There is a subtlety here regarding the transformation $V$. If $\mathcal{C}_i\neq 0$, $V$ contains topological singularities, which changes the topological properties under the transformation. As a consequence, the winding number $\nu$ is in general not well defined. We illustrate this point by a simple example. Consider quantum quench from $\mathcal{C}_i=1$ (ground state of model (7) with $q=1$) with post-quench Hamiltonian $h=\sigma_z$. For simplicity, we set $M\lesssim 2$. $\Gamma$ is then a circle $k_x^2+k_y^2=2(2-M)\equiv r^2$ and parameterized by $\theta$, $k_x=r\cos\theta$, $k_y=r\sin\theta$. The pre-quench Hamiltonian along $\Gamma$ is $h_0=\cos\theta\sigma_x+\sin\theta\sigma_y$. The two eigenstates of $h_0$ are
\begin{eqnarray}
|\xi_0^{+}\rangle=\frac{1}{\sqrt{2}}\left(\begin{array}{c}
1\\
e^{i\theta}\\
\end{array}
\right);~~~~|\xi_0^{-}\rangle=\frac{1}{\sqrt{2}}\left(\begin{array}{c}
e^{-i\theta}\\
-1\\
\end{array}
\right).
\end{eqnarray}
There exist many different choices for the unitary transformation $V$ to diagonalize $h_0(\theta)$, corresponding to different gauge choices of the above eigenstates. We consider the following two transformations,
\begin{eqnarray}
V_1=\frac{1}{\sqrt{2}}\left(\begin{array}{cc}
1 & e^{-i\theta}\\
e^{i\theta} & -1\\
\end{array}
\right);~~~~V_2=\frac{1}{\sqrt{2}}\left(\begin{array}{cc}
e^{-i\theta} & 1\\
1 & -e^{i\theta}\\
\end{array}
\right).
\end{eqnarray}
Obviously $V_1^{\dag}h_0V_1=V_2^{\dag}h_0V_2=\sigma_z$. However, $V_1^{\dag}U_lV_1=i(\sin\theta\sigma_x-\cos \theta\sigma_y)$, and $V_2^{\dag}U_lV_2=i(-\sin\theta\sigma_x-\cos \theta\sigma_y)$. Their tilting phases have opposite signs: $\chi_1=-\chi_2=\theta-\frac{\pi}{2}$, giving rise to opposite winding numbers $\nu_1=-\nu_2=1$. Finally we note that although the winding topology of $\Gamma$ relies on the triviality of $V$, our characterization from the $\pi$-defect and its topological charge is still valid.

\subsection{VI. Dynamical topology of Bernevig-Hughes-Zhang model}
In this section, we show the framework of loop unitary can be applied to characterize the quench dynamics of multi-band models from different symmetry classes. As an example, we focus on the Bernevig-Hughes-Zhang (BHZ) model \cite{bhz1}, which describes a four-band topological insulator belonging to class AII. Its bulk topology is characterized by a $\mathbb{Z}_2$ invariant $\nu_2$ and manifests via the appearance of helical edge states on the boundary. We consider a lattice version \cite{bhz2} of the BHZ model as follows (assume $M>0$): 
\begin{eqnarray}
H_{BHZ}(\bm k)=\sin k_x\tau_y\otimes\sigma_0+\sin k_y\tau_x\otimes\sigma_z+(M-\cos k_x-\cos k_y)\tau_z\otimes\sigma_0.
\end{eqnarray}
The band structure is given by $\pm E_{\bm k}=\pm\sqrt{\sin^2 k_x+\sin^2 k_y+(M-\cos k_x-\cos k_y)^2}$, each branch is two-fold degenerate. It is easy to check the above Hamiltonian satisfies time-reversal symmetry $\Theta=\tau_0\otimes i\sigma_yK$ with $\Theta H_{BHZ}(\bm k)\Theta^{-1}=H_{BHZ}(-\bm k)$ as well as inversion symmetry $I=\tau_z\otimes\sigma_0$ with $I H_{BHZ}(\bm k)I^{-1}=H_{BHZ}(-\bm k)$. The $\mathbb{Z}_2$ invariant $\nu_2$ can be calculated from the parities \cite{bhz2} of the Bloch states at the four time-reversal invariant momenta $(0,0)$, $(0,\pi)$, $(\pi,0)$ and $(\pi,\pi)$. For $M>2$, $\nu_2=0$, the system is a trivial insulator; while for $0<M<2$, $\nu_2=1$, the system is a topological insulator due to band inversion and supports helical edge states. 

Let us consider a quantum quench $H_0\rightarrow H_{BHZ}$ at $t=0$, with pre-quench Hamiltonian $H_0=\tau_z\otimes\sigma_0$. The loop unitary operator reads $U_l(t,\bm k)=e^{-i h_{BHZ}(\bm k) t}e^{i H_0 t}$, where $h_{BHZ}(\bm k)$ is the rescaled post-quench Hamiltonian $h_{BHZ}(\bm k)=\frac{H_{BHZ}(\bm k)}{E_{\bm k}}$. The loop unitary is time-periodic $U_l(t+\pi,\bm k)=U_l(t,\bm k)$ and defines a mapping from the momentum-time three-torus $\textrm{T}^3$ to $U(4)$ matrices. 

Now let us construct the dynamical invariant. A key observation is that the 3-winding number $W_3[U_l]$ (Eq. (3) in the main text) is always zero because the contributions from the Kramers pairs cancel each other \cite{z2dynamics}. The loop unitary satisfies time-reversal symmetry $\Theta U_l(t,\bm k)\Theta^{-1}=U_l(-t,-\bm k)$. We denote the desired $\mathbb{Z}_2$-valued dynamical invariant as $\delta$, which can be obtained either by following the interpolation method \cite{wangzhong} or by using a contracted evolution operator \cite{z2dynamics}. We take the second route as example. Formally, instead of calculating the $W_3[U_l]$, we consider the following smooth map $V_l(t,\bm k)$ from $\textrm{T}^3$ to $U(4)$ such that
\begin{eqnarray}
&&V_l(t,\bm k)=U_l(t, \bm k)~~~~~~~~~~~~~~~~\text{for}~~~ 0<t<\pi/2\notag\\
&&\Theta V_l(t,\bm k)\Theta^{-1}=V_l(t,-\bm k)~~~~~~\text{for}~~~\pi/2<t<\pi\notag\\
&&V_l(0,\bm k)=V_l(\pi,\bm k)=\textrm{I}.
\end{eqnarray}
The dynamical invariant is defined by the 3-winding number of the contracted loop unitary $V_l$ modulo 2,
\begin{eqnarray}
\delta=W_3[V_l] \bmod 2.
\end{eqnarray}
We note the $\mathbb{Z}_2$ invariant $\delta$ is well defined. The existence of the contracted loop unitary, and the independence of $\delta$ on the choice of contraction have been proved \cite{z2dynamics}. For our quench problem, if the post-quench Hamiltonian is topological ($0<M<2$), $\nu_2=1$, the dynamical topology is then characterized by $\delta=1$, with the emergence of $\pi$-defects in the phase band of $U_l$; however, if the post-quench Hamiltonian is trivial ($M>2$), the dynamical invariant is $\delta=0$ without any $\pi$-defect in the phase band. 

Thus our framework proves fruitful for the example of four-band topological insulators in AII class. For more complicated cases and other symmetry classes, the construction of dynamical invariant from the loop unitary operator may require taking into account additional constraints \cite{roy,wangzhong}. Recently, we have studied the quench dynamics of a 3D Hopf insulator and constructed the $\mathbb{Z}_2$ dynamical invariant using the loop unitary \cite{hopfquench}.\newline

\subsection{VII. Experimental detection of DQPT curve and topological charge $W_3$}
In this section, we discuss how to experimentally determine the DQPT curve and topological charge of $\pi$-defect via time- and momentum-resolved Bloch-state tomography \cite{azi1,azi2,tomograph1,tomograph2}. As mentioned in the main text, the DQPT curve can be obtained by tracing the emergent dynamical vortices. Here we take the post-quench Hamiltonian $H_{q=1}$ (Eq. (7) in the main text) with pre-quench Hamiltonian $H_0=\sigma_z$ as an example. 

The Bloch-state tomography is based on a double quench protocol \cite{azi1}, which involves a projection of the time-dependent quantum state (e.g. the time-evolved state $|\xi(\bm k,t)\rangle$ after the quench) onto a target tomography Hamiltonian. The target Hamiltonian can be chosen as $H_{tomo}=\frac{\Delta}{2}\sigma_z$, with $\Delta$ the band gap. This method allows mapping of the full quantum-mechanical wavefunction at any time after the quantum quench. The key idea of tomography is to observe the ensuing momentum-dependent precession (after the projection) on the Bloch sphere, which translates into a Bloch oscillation after time-of-flight expansion. We denote the precession time as $t^{prec}$. Formally, for a two-component state $|\xi(\bm k,t)\rangle=\left(\begin{array}{cc}\cos\frac{\theta_{\bm k,t}}{2}\notag\\ \sin\frac{\theta_{\bm k,t}}{2}e^{i\phi_{\bm k,t}}\end{array}\right)$, the momentum distribution after the time-of-flight measurement reads \cite{azi1,quenchexp2,quenchexp1}
\begin{eqnarray}\label{density}
n(\bm k,t^{prec})=f(\bm k)[1+\sin\theta_{\bm k,t}\cos(\Delta~t^{prec}+\phi_{\bm k,t})].
\end{eqnarray}
Here $f(\bm k)$ is the Fourier transformed Wannier function. From Eq. (\ref{density}), we can clearly see $\theta_{\bm k,t}$, $\phi_{\bm k,t}$ can be extracted from the oscillatory signal of momentum distributions. 
\begin{figure}[t]
\includegraphics[width=5.5in]{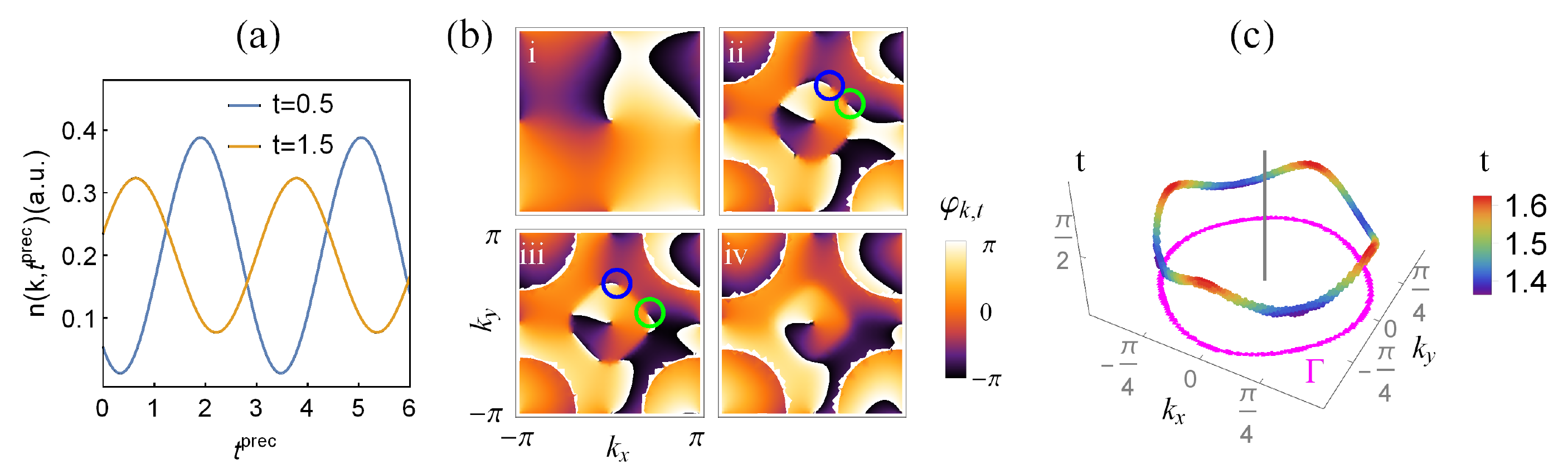}
\caption{Time-resolved Bloch-state tomography of the post-quench state $|\xi(\bm k,t)\rangle$. (a) Density oscillations with respect to the precession time $t^{prec}$ for the double quench protocol, from which one can obtain the azimuthal phase $\phi_{\bm k,t}$ and amplitude $\theta_{\bm k,t}$. The momentum is chosen as $\bm k=(\frac{\pi}{3},\frac{\pi}{3})$. (b) Instantaneous phase profile $\phi_{\bm k,t}$ in the first BZ for four different evolution times. From i to iv, $t=0.5$, $t=1.45$, $t=1.55$, and $t=1.8$. The static vortices at the high-symmetry points remain for all evolution times. The emergent dynamical vortex and anti-vortex pair are marked by blue and green circles, respectively, with the vorticity defined in Ref. \cite{quenchexp2}. For clarity, only one pair are marked. There are four pairs in total, symmetrically distributed around $\bm k=(0,0)$. (c) The trajectory of the dynamical vortices in the momentum-time space. The projected magenta curve in the 2D BZ is the DQPT curve $\Gamma$. The vertical line marks the static vortex located at $\bm k=(0,0)$. In all figures, the post-quench Hamiltonian is $H_{q=1}$ (Eq. (7) in the main text) with $M=1.2$.}
\label{figs2}
\end{figure}

The numerical simulations for the quench under consideration are illustrated in Fig. \ref{figs2}. The initial state $|\xi_0\rangle$ lies at the south pole on the Bloch sphere for all momentum modes. Fig. \ref{figs2}(a) depicts the density oscillations for post-quench states $|\xi(\bm k,t=0.5)\rangle$ and $|\xi(\bm k,t=1.5)\rangle$ at a representative momentum $\bm k=(\frac{\pi}{3},\frac{\pi}{3})$. From the oscillation, we can extract their corresponding $\theta$ and $\phi$. Fig. \ref{figs2}(b) depicts the momentum-resolved phase profile $\phi_{\bm k,t}$ for quench time $t=0.5$, $t=1.45$, $t=1.55$, and $t=1.8$ obtained from their density oscillations. After the quench, the time-evolved state will spread on the Bloch sphere. Once one momentum mode reaches the north pole, the neighbouring modes will occupy a patch around the pole, resulting in a winding pattern of the azimuthal phase $\phi$. Reaching the north pole corresponds to the emergence of a dynamical vortex-anti-vortex pair in the Brillouin zone (BZ). Our numerics show the vortex pair appear at $t\approx1.388$ and annihilate each other at $t\approx1.603$. In additional to the dynamical vortices, there exist static vortices at the four high-symmetry points $\bm k=(0,0)$, $(0,\pi)$, $(\pi,0)$, and $(\pi,\pi)$. These static vortices exist for all evolution times and stay fixed during the time evolution. We note that DQPT occurs when the time-evolved state reaches the north pole. Hence by tracing the dynamical vortices, we can directly obtain the DQPT curve $\Gamma$, which is nothing but the projection of the trajectories of dynamical vortices onto the 2D BZ, as shown in Fig. \ref{figs2}(c). The windings along $\Gamma$ is readily obtained using the same tomography data. 
\begin{figure}[h]
\includegraphics[width=3.8in]{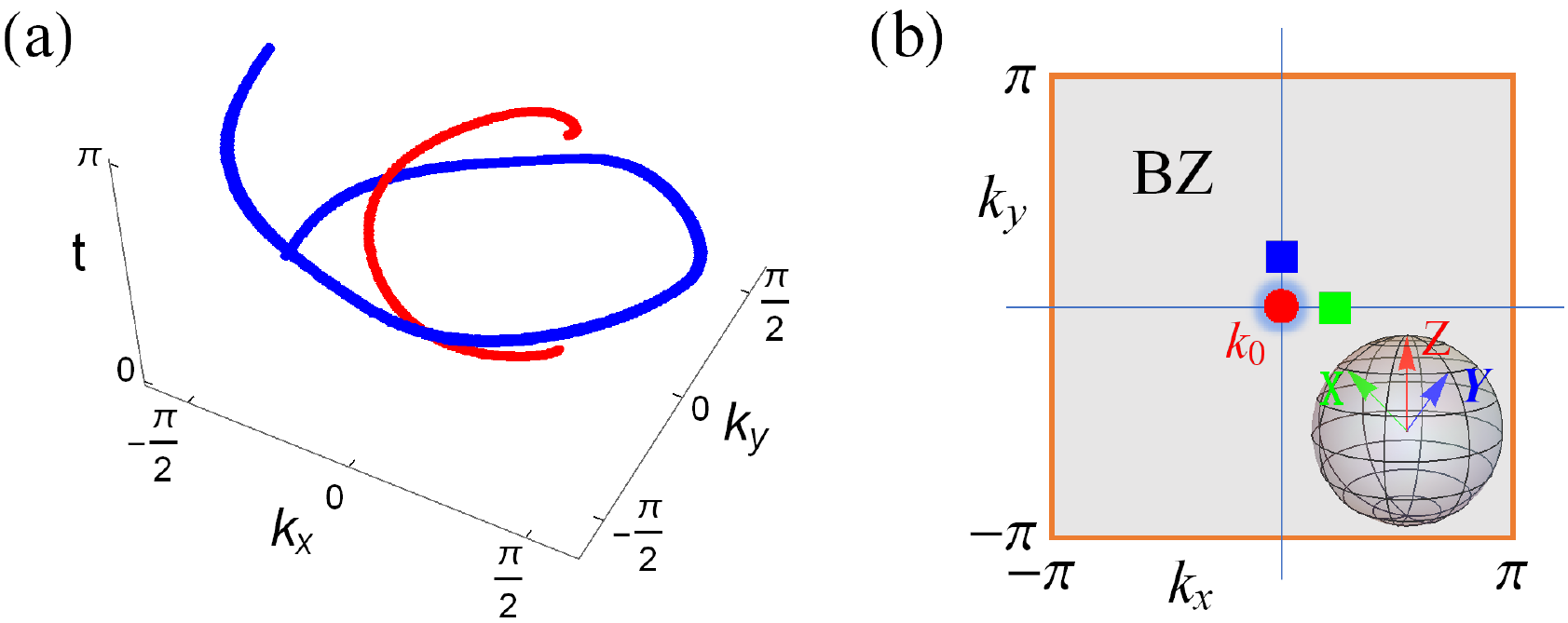}
\caption{Quantum quench starting from nontrivial initial state with pre-quench Hamiltonian $H_{q=1}$, $M=1.2$. (a) Preimages of two chosen quantum states $(\frac{1}{2},\frac{\sqrt{3}}{2})^T$ (blue) and $(\frac{\sqrt{3}}{2},-\frac{1}{2})^T$ (red) in the momentum-time space. (b) Tomography of the two nearby points $\bm k_1=(\delta k,0)$ (green) and $\bm k_2=(0,\delta k)$ (blue) on the $t=\frac{\pi}{2}$ plane, $\delta k$ is set as $0.2\pi$. The $\pi$-defect is located at $\bm k_0=(0,0)$ (red dot). The inset shows the corresponding Bloch vectors $X$ (green), $Y$ (blue), $Z$ (red) of $\xi(\bm k_1,\pi/2)$, $\xi(\bm k_2,\pi/2)$ and $\xi(\bm k_0,\pi/2)$ on the Bloch sphere. Here $XYZ$ is right-handed, yielding topological charge $Q=-1$.}
\label{figs3}
\end{figure}

Based on the Bloch-state tomography, we can further determine the Hopf linking number $\mathcal{L}$ \cite{quenchexp2} from the chirality of the vortex contour if the initial state is trivial. For the case considered above, the dynamical vortices constitute the preimages of $(1,0)^T$. $\mathcal{L}$ is nothing but the linking number of the static and dynamical vortices \cite{qq5} [See Fig. \ref{figs2}(c)], $\mathcal{L}=1$. For quantum quenches starting from a nontrivial initial state, however the Hopf linking number is not well defined. To characterize the dynamical topology, we have to resort to the loop unitary and its associated $\pi$-defects. As discussed in the main text, the dynamical invariant is the total charges of these $\pi$-defects $W_3=\sum_i Q_i$. In the following we show how to determine the topological charges using the Bloch-state tomography, which can identify $|\xi(\bm k,t)\rangle$ for any $\bm k$ and $t$. 

The procedure contains two steps. (\textit{i}) First we need to identify the location of the $\pi$-defects in the BZ. This is simple because at the $\pi$-defects, the post-quench Hamiltonian vector $\hat{\bm h}$ is parallel to the initial spin vector $\hat{\bm n}_0$, i.e., $\hat{\bm h}=\hat{\bm n}_0$. The time-evolved state $|\xi(\bm k,t)\rangle$ will stay fixed (time-independent) on the Bloch sphere after the quench. (\textit{ii}) The second step is to measure the topological charge $Q_i$ for each defect. To linear order, the loop unitary takes $U_l(\delta k_x,\delta k_y,\delta t)=-\textrm{I}-i\delta k_i K_{ij}\sigma_j$ with $\delta k_i\in(\delta k_x,\delta k_y,\delta t)$ being deviations from the defect. The topological charge is given by $Q_i=\sgn(\det(K))$ \cite{phaseband}. Experimentally, the $K$ matrix can be determined from the tomography of the nearby points in the BZ \cite{chargeexp}. 

Now we demonstrate the above procedure by an example with pre-quench Hamiltonian $H_{q=1}$ and post-quench Hamiltonian $H=\sigma_z$. The initial state is the ground state of $H_{q=1}$ with Chern number $\mathcal{C}_i=1$. The main results are summarized in Fig. \ref{figs3}. In contrast to the previous case, the preimages of two chosen quantum states form open helix link in the momentum-time space, as shown in Fig. \ref{figs3}(a) and the Hopf invariant is not well defined. The $\pi$-defect is located at $\bm k_0=(0,0)$ and the initial state corresponds to the north pole on the Bloch sphere. The density profile shows no oscillation at $\bm k_0$. Due to the simple form of post-quench Hamiltonian, the topological charge $Q$ can be extracted from tomography of the two nearby points along the $k_x$ and $k_y$ axis, as shown in Fig. \ref{figs3}(b). Ideally, the nearby points should be infinitesimally away from the defect. However, in real experiments, a finite distance is sufficient due to the quantization and stability of the charge. The Bloch vectors of the corresponding time-evolved state $\xi(\bm k,t=\frac{\pi}{2})$ determined from the tomography are depicted in Fig. \ref{figs3}(b). By labeling the three vectors as $X$, $Y$, $Z$, respectively, the topological charge is then the chirality of these vectors. If $XYZ$ is right-handed, the topological charge is $-1$; if $XYZ$ is left-handed, the topological charge is $+1$. For our case, the charge is $Q=-1$, consistent with the Chern number change across the quantum quench $W_3=\mathcal{C}_f-\mathcal{C}_i=0-1=-1$. 

\end{document}